\documentclass[11pt]{article}

\usepackage{natbib}
\usepackage{fullpage}
\usepackage{epsfig}
\usepackage[framemethod=TikZ]{mdframed}

\usepackage{tikz} 
\usetikzlibrary{shapes.misc,shadows}

\usepackage{graphicx} 

\mdfsetup{frametitlealignment=\centering}
\mdfdefinestyle{offset}{backgroundcolor=white,linecolor=black,innerrightmargin=15pt,innermargin=10pt,outermargin=10pt,innertopmargin=.5\baselineskip,innerbottommargin=.5\baselineskip}

\usepackage{amsmath}
\newtheorem{theorem}{Theorem}[section]
\usepackage{natbib}
\usepackage{xspace}

\newcommand{\dist}{F}

\newcommand{\disti}[1][i]{{\dist_{#1}}}

\newcommand{\val}{v}

\newcommand{\vali}[1][i]{{\val_{#1}}}

\newcommand{\vv}{\varphi}
\newcommand{\vvi}[1][i]{{\vv_{#1}}}

\newcommand{\dens}{f}

\newcommand{\densi}[1][i]{{\dens_{#1}}}

\newcommand{\bid}{b}

\newcommand{\bidi}[1][i]{{\bid_{#1}}}

\newcommand{\argmax}{\operatornamewithlimits{argmax}}

\newcommand{\eps}{\epsilon}

\usepackage{enumitem}

\begin{document}

\title{Approximately Optimal Mechanism Design
(Preprint)\thanks{January 1, 2019. To appear in {\em Annual Reviews of Economics},
  August 2019.
    Based in part on an article by \cite{Rou14}.}}

\author{Tim Roughgarden\thanks{Computer Science Department, Stanford University, Stanford, CA, 94305; email: tim@cs.stanford.edu}
and Inbal Talgam-Cohen\thanks{Computer Science Department, Technion -- Israel Insitute of Technology, Haifa, Israel, 3200003; email: italgam@cs.technion.ac.il}
}

\date{}

\maketitle

\section{The Optimal Mechanism Design Paradigm: Success Stories}
\label{sec:opt-paradigm}

Optimal mechanism design enjoys a beautiful and well-developed theory,
and also a number of killer applications. Let's review two famous
examples.

\subsection{Example: The Vickrey Auction}\label{sub:vickrey}

In the Vickrey or second-price single-item auction~\citep{V61}, there
is a single seller with a single item; assume for simplicity that the
seller has no value for the item.  There are $n$ bidders, and each
bidder~$i$ has a valuation $\vali$ that is unknown to the seller.
The Vickrey auction is designed to maximize the welfare, which in a
single-item auction just means awarding the item to the bidder with
the highest valuation.  This sealed-bid auction collects a bid from
each bidder, awards the item to the highest bidder, and charges the
second-highest price.  The point of the pricing rule is to ensure that
truthful bidding is a dominant strategy for every bidder.  Provided
every bidder follows its dominant strategy, the auction maximizes
welfare ex post (that is, for every valuation profile).

In addition to being theoretically optimal, the Vickrey auction has a
simple and appealing format.  Plenty of real-world examples resemble
the Vickrey auction.  In light of this confluence of theory and
practice, what else could we ask for?  To foreshadow what lies ahead,
we mention that when selling multiple non-identical items, the
generalization of the Vickrey auction is much more complex.

\subsection{Example: Myerson's Auction}\label{ss:iid}

What if we want to maximize the seller's revenue rather than the
social welfare?  Since there is no single auction that maximizes
revenue ex post, the standard approach here is to maximize the
expected revenue with respect to a prior distribution over bidders'
valuations.  Assume bidder $i$'s valuation is drawn independently from a
distribution $\disti$ that is known to the seller.  For the moment,
assume also that bidders are homogeneous, meaning that their
valuations are drawn i.i.d.\ from a known distribution $\dist$.

\cite{Mye81} identified the optimal auction in this context, which
is a simple twist on the Vickrey auction --- a
second-price auction with a reserve price~$r$.\footnote{That is, the
  winner is the highest bidder with bid at least $r$, if any.  If
  there is a winner, it pays either the reserve price or the
  second-highest bid, whichever is larger.}  Moreover, the optimal
reserve price is simple and intuitive---it is the {\em monopoly
  price} $\argmax_p [p \cdot (1-\dist(p))]$ for the distribution $F$,
the optimal take-it-or-leave-it offer to a single
bidder with valuation drawn from $\dist$.  Thus, to implement the
optimal auction, you don't need to know much about the valuation
distribution $\dist$---just a single statistic, its monopoly price.

Once again, in addition to being theoretically optimal, Myerson's
auction is simple and appealing.  It is more or less equivalent to an
eBay auction, where the reserve price is implemented using an opening
bid.  Given this success, why do we need to enrich the traditional
optimal mechanism design paradigm?  As we'll see, when bidders'
valuations are not i.i.d., the theoretically optimal auction is much
more complex and no longer resembles the auction formats that are
common in practice.

\subsection{The Optimal Mechanism Design Paradigm}

Having reviewed two well-known examples, let's zoom out and be more
precise about the optimal mechanism design paradigm.  The first step
is to identify the design space of possible mechanisms, such as the
set of all sealed-bid auctions.  The second step is to specify some
desired properties.  In this survey, we focus only on cases where the
goal is to optimize some objective function that has cardinal meaning,
and for which relative approximation makes sense.  We have in mind
objectives such as the seller's revenue (in expectation with respect
to a prior) or social welfare (ex post) in a transferable utility
setting.  The goal of the analyst is then to identify one or all
points in the design space that possess the desired properties---for
example, to characterize the mechanism that maximizes the welfare or
expected revenue.

What can we hope to learn by applying this framework?  The traditional
answer is that by solving for the optimal mechanism, we hope to
receive some guidance about how to solve the problem.  With the
Vickrey and Myerson auctions, we can take the theory quite literally
and simply implement the mechanism advocated by the theory.
More generally, one looks for features present in the theoretically
optimal mechanism that seem broadly useful.  For example, Myerson's
auction suggests that combining welfare maximization with suitable
reserve prices is a potent approach to revenue-maximization.

There is a second, non-traditional answer that we exploit
explicitly when we extend the paradigm to accommodate approximation.
Even when the theoretically optimal mechanism is not directly useful
to the practitioner, for example because it is too complex, it is
directly useful to the analyst.  The reason is that the performance
of the optimal mechanism can serve as a benchmark, a yardstick against
which we measure the performance of other designs that stand a chance
of being implemented.

\section{The Optimal Mechanism Design Paradigm: Failure Modes}
\label{sec:failure-modes}

The Vickrey and Myerson auctions are exceptions that prove a rule:
theoretically optimal mechanisms are generally too complex to be used
in practice.  ``Complexity'' can take many forms, including excessive
computation, excessive communication, or unrealistic informational
assumptions.  We next illustrate this point with three
examples.  These examples motivate the alternatives to optimal
mechanisms described in Sections~\ref{sec:case-study1}--\ref{sec:case-study3}.

\subsection{Optimal Single-Item Auctions (Excessive Information)}\label{sub:fail1}

We now return to expected revenue-maximization in single-item
auctions, but allow {\em heterogeneous} bidders, meaning that
each bidder $i$'s private valuation $\vali$ is drawn independently
from a distribution $\disti$ that is known to the seller.
\cite{Mye81} characterized the optimal auction, as a function of
the distributions $\dist_1,\ldots,\dist_n$.
We assume for simplicity that each distribution $\disti$ has bounded
support and a density function $\densi$.

The trickiest step of Myerson's optimal auction is the first one,
where each bid $\bidi$ is transformed into a {\em virtual bid}
$\vvi(\bidi)$, defined by
\[
\vvi(\bidi) = \bidi - \frac{1-\disti(\bidi)}{\densi(\bidi)}.
\]
The exact functional form in this equation is not important for this
survey, except to notice that computing $\vvi(\bidi)$ requires
knowledge of the distribution, namely of $\densi(\bidi)$ and $\disti(\bidi)$.

Given this transformation, the rest of the auction is
straightforward.  The winner is the bidder with the highest positive
virtual bid (if any).  To make truthful bidding a dominant strategy,
the winner is charged the minimum bid at which it would continue to be
the winner.\footnote{We have only described the optimal auction in the
  special case where each distribution $\disti$ is {\em regular},
  meaning that the virtual valuation functions $\vvi$ are
  nondecreasing.  The general case ``monotonizes'' or ``irons'' the virtual
  valuation functions
and then applies the same three steps~\citep{Mye81}.
(Monotonicity is essential for incentive-compatibility.)}

When all the distributions $\disti$ are equal to a common~$\dist$, and
hence all virtual valuation functions $\vvi$ are identical, the
optimal auction simplifies and is simply a second-price auction with a
reserve price of $\vv^{-1}(0)$, which turns out to be the monopoly
price for $\dist$.  In this special case, the optimal auction requires
only modest distributional knowledge (the monopoly price).  In
general, the optimal auction does not simplify further than the
description above.  A major impediment to implementing such a
``virtual welfare maximizer'' is that accurate distributional details
are not always available -- this widely-accepted criticism is known as
Wilson's doctrine \citep{Wil87}. 
Second, even if such details are
available, the corresponding optimal mechanism can be too inscrutable for
real-world deployment.  For example, on the second point, an
optimal single-item auction might award the item to a low bidder
over a high bidder (even if the latter clears its reserve).

\subsection{Welfare-Maximizing Multi-Item Auctions (Excessive Communication)}\label{sub:fail2}

In the standard setup for allocating multiple items via a
combinatorial auction, there are~$n$ bidders and $m$ non-identical
items.  Each bidder has, in principle, a different private valuation
$\vali(S)$ for each bundle~$S$ of items it might receive.  Thus,
each bidder has $2^m$ private parameters.  In this example, we assume
that the objective is to determine an allocation $S_1,\ldots,S_n$ that
maximizes the social welfare $\sum_{i=1}^n \vali(S_i)$.

The Vickrey auction can be extended to the case of multiple items;
this extension is the Vickrey-Clarke-Groves (VCG)
mechanism~\citep{V61,C71,G73}.  The VCG mechanism is a
direct-revelation mechanism, so each bidder $i$ reports a valuation
$\bidi(S)$ for each bundle of items~$S$.  The mechanism then computes an
allocation that maximizes welfare with respect to the reported
valuations.  As in the Vickrey auction, suitable payments make truthful
revelation a dominant strategy for every bidder.

Even with a small number of items, the VCG mechanism is a non-starter
in practice, for a number of reasons~\citep{AM06}.  For example, the
VCG mechanism, as a direct-revelation mechanism, solicits $2^m$
numbers from each bidder.  This is an exorbitant number: roughly a
thousand parameters already when $m=10$, roughly a million when
$m=20$.  
In modern spectrum auctions, $m$ might be in the hundreds or
larger.\footnote{\cite{cramton} writes: ``The setting of spectrum
  auctions is too complex to guarantee full efficiency.''
\label{foot:cramton}}

\subsection{Welfare-Maximization with Single-Minded Bidders (Excessive
  Computation)}\label{sub:fail3}

If bidders' preferences are easy to communicate, does the VCG mechanism
become easy to implement?
For example, suppose each
bidder~$i$ is {\em single-minded} and only cares about a (publicly
known) subset $T_i$ of items \citep{LOS02}.
Bidder~$i$ has a private value $\vali$
for every superset of $T_i$, and~0 for every other set.
This is a single-parameter environment, so communication between the
bidders and the mechanism is not an issue.
``All'' the VCG mechanism has to do is compute a
welfare-maximizing allocation (with respect to the reported
valuations) and appropriate prices.

The problem is that, for single-minded bidders and many other examples
of succinctly described valuations, it is difficult to compute a
welfare-maximizing allocation in a reasonable amount of time (less
than a year, say).  The problem is that the number of candidate
solutions grows {\em exponentially} with the number~$n$ of bidders.  A
subset~$W$ of bidders can all receive their desired subsets
simultaneously if and only if if $T_i \cap T_j = \emptyset$ for
distinct~$i,j \in W$ (since no item can be allocated more than once).
With~$n$ bidders, there are~$2^n$ possibilities for~$W$.  For modestly
large~$n$ (at least~50, say), there is no hope of checking them all in
a reasonable amount of time.\footnote{Auctions for online advertising
  can have dozens or even hundreds of participants.  The reverse
  auction in the FCC Incentive Auction
  (Section~\ref{sub:motivating-ex}) had thousands of participants.}
For some computational problems with exponentially many candidate
solutions, there is a clever algorithm that shortcuts to the optimal
solution while examining only a tiny fraction of the possibilities.
For ``$NP$-hard'' optimization problems, including the problem of
welfare-maximization with single-minded bidders, the exponential
scaling appears fundamental, with no clever shortcut in sight.

\section{Approximately Optimal Mechanism Design}
\label{sec:approx-paradigm}

\subsection{Benchmarks and Approximate Optimality}

The examples in Section~\ref{sec:failure-modes} demonstrate that, for
many different reasons, it is not always feasible to implement the
theoretically optimal mechanism.  To give better design guidance in
such settings, we have no choice but to take a different approach.
This brings us to the main theme of this survey: using the
relaxed goal of {\em approximate optimality} to make new progress on
fundamental but challenging mechanism design problems.

To study approximately optimal mechanisms, we again begin with a
design space and an objective function.  Often the design space will
proxy for the set of ``plausibly implementable mechanisms,'' and is
accordingly limited by side constraints such as a ``simplicity''
constraint.  For example, we later consider mechanisms with a
restricted number of pricing parameters, with low-dimensional bid
spaces, and with limited computational power.

The new ingredient of the paradigm is a {\em benchmark}.  This is a
target objective function value that we would be ecstatic to
achieve.  Generally, the working hypothesis will be that no mechanism
in the design space realizes the full value of the benchmark, so the
goal is to get as close to it as possible.
In the examples we discuss, where the design space is limited by a
simplicity constraint, a natural benchmark is the performance
achieved by an unconstrained, arbitrarily complex mechanism.
The goal of the analyst is to identify a mechanism in the design space
that approximates the benchmark as closely as possible.

A typical positive result in approximately optimal mechanism design
identifies a mechanism in the desired design space that always
guarantees an objective function value (social welfare, expected
revenue, etc.) that is at least an $\alpha$ percentage of the benchmark
value.  (The closer $\alpha$ is to~100\%, the better.)  A typical negative
result proves that there is no mechanism in the design space with
such a guarantee (for some fixed percentage $\alpha$).

\subsection{Goals of Approximately Optimal Mechanism Design}\label{sub:goals}

What is the point of applying this design paradigm?  The first goal is
exactly the same as with the traditional optimal mechanism design
paradigm.  Whenever you have a principled way of choosing one
mechanism from many, you can hope that the distinguished mechanism is
literally useful or highlights features that are essential to
good designs.  The approximation paradigm provides a novel way to
identify candidate mechanisms.

There is a second reason to use the approximately optimal mechanism
design paradigm, which has no analog in the traditional approach.
The approximation framework allows the analyst to quantify the cost
of imposing side constraints on a mechanism design space.  For
example, if there is a simple mechanism with performance close to
that of the best arbitrarily complex mechanism, then this fact suggests
that simple solutions might be good enough.  Conversely, if every
point in the design space is far from the benchmark, then this
provides a forceful argument that complexity is an essential feature
of every reasonable solution to the problem.  Our second case study 
(Section~\ref{sec:case-study2}) is a particularly clear example of this perspective.

\subsection{Coming Up: Three Case Studies}

Sections~\ref{sec:case-study1}--\ref{sec:case-study3} describe three
such instantiations, each addressing a different drawback of
theoretically optimal mechanisms.  First, we study expected
revenue-maximization in single-item auctions, with bidders that have
independent but not necessarily identically distributed valuations.
Virtual welfare maximizers are an overparameterized class of auctions,
and selecting the right one requires detailed distributional
knowledge.  We use the approximation paradigm to understand fundamental
trade-offs between optimality and simplicity.

Our second case study concerns the problem of selling multiple
non-identical items to maximize the social welfare.  The theoretically
optimal mechanism is well known (the VCG mechanism) but suffers from
several drawbacks that preclude direct use.  We apply the
approximation paradigm to identify when mechanisms with
low-dimensional bid spaces can perform well, and when high-dimensional
bid spaces are necessary for non-trivial welfare guarantees.

Our final case study concerns settings where computation is the
primary obstacle to optimality.  Multi-unit
auctions are one canonical example.  We use the approximation
framework to identify mechanisms that guarantee near-optimal social
welfare and are also computationally efficient.

An enormous amount of research over the past twenty years, largely but
not entirely in the computer science literature, can be viewed as
instantiations of the approximately optimal mechanism design
paradigm.  The case studies in this survey are representative but far
from exhaustive.  The book of \cite{Har17} is a good source for
additional examples.

\subsection{In Defense of Approximation Ratios}

The positive results in our three case studies have the form:
 ``the objective function value of the simple mechanism $M$ is always
at least an $\alpha$ percentage of that of the (complex) optimal
mechanism.''  In some cases, $\alpha$ will be close to 100\%, and the
utility of the guarantee is self-evident.  In most settings, however,
the best-possible approximation guarantee is bounded away from 100\%.
What use is an approximate guarantee of, say, 63\%?\footnote{For the
  most part, we focus on relative approximation guarantees, which have
  the advantage of canceling out units of measurement.  
Absolute approximation guarantees are also meaningful in some settings
(e.g., Theorem~\ref{t:mr15}).}

In the authors' experience, researchers tend to fixate
unduly on and take too literally the numerical values in approximation
guarantees.  There are several points to keep in mind:
\begin{enumerate}

\item Both of the primary motivations for applying the approximately
  optimal mechanism design paradigm (Section~\ref{sub:goals}) strive
  for qualitative rather than quantitative insights. This holds both
  for identifying mechanism features that are potentially useful in
  practice, and for assessing the cost of a simplicity side-constraint
  on the mechanism.  For example, if the pursuit of a best-possible
  approximation guarantee justifies a widely-used mechanism or guides
  the analyst to an interesting new mechanism, is the exact numerical
  value of the guarantee so important?

\item A reader who, against our advice, insists on interpreting
  approximation guarantees literally, is likely to ask: ``what about
  the other 37\% of the welfare or expected revenue being left on the
  table?''  But in all of the canonical applications of approximately
  optimal mechanism design, the benchmark of full optimality is 
only a utopia in the analyst's mind, and not
  one of the available options.  For example, in a multi-item auction
  with more than a few items, it is flat-out impossible to implement a
  welfare-maximizing mechanism like the VCG mechanism.
  The choice is not whether to implement an optimal mechanism; it's
  whether to implement a suboptimal mechanism that has a good
  approximation guarantee or one that doesn't.  While the mechanism
  with the best-possible approximation guarantee may or may not be the
  best one to implement in practice, it is always worth considering.

\item Approximation guarantees are usually ``worst case,'' meaning
  that they hold for every possible setting (e.g., for an arbitrary
  valuation profile, or in expectation for an arbitrary prior
  distribution).  An approximately optimal mechanism usually performs
  better than its worst-case guarantee in most settings of interest.
  For example, a mechanism with a worst-case guarantee of 50\% might
  well achieve at least 90\% of the benchmark value on ``typical''
  inputs.  In some cases, this property can be proved formally by
  establishing better approximation guarantees under additional
  assumptions about the setting; in other cases, the argument is best
  made through simulations.

\item Is a number like ``63\%'' big or small?  As in real life, the
  answer depends on the context.\footnote{A professional basketball
    team that wins 63\% of its games is good but not great, while a
    baseball team with the same record would be one of the favorites
    to win the World Series.  Similarly, is a six-week turnaround for
    referee reports fast or slow?  The answer depends on whether the
    submission was sent to {\em Econometrica} or {\em Science}.}  The
  best way to justify theoretically an approximation guarantee is to
  prove a matching impossibility result, stating that there is no
  mechanism in the class of interest with a superior guarantee.  With
  a few exceptions, like ``large market''-type results
  (Section~\ref{sub:large-markets}), optimal approximation guarantees
  tend to be bounded well away from 100\% (e.g., 50\% or 63\%).

\end{enumerate}
Whatever the merits of approximately optimal mechanism design, the
unimplementability of optimal mechanisms in complex settings is real
and will not go away (cf., footnotes~\ref{foot:cramton} and~\ref{foot:milgrom}). Any theorist
who wants to reason seriously about such settings must work with an
alternative to the classical optimal mechanism design paradigm.
Approximation is by no means the only possible alternative (see also
Section~\ref{sec:alternatives}), but it is one of the most successful
approaches to date.

Finally, like any general analysis framework, the approximation
paradigm can be abused and should be applied with good taste.  In
settings in which the paradigm does not give meaningful results,
different modeling or benchmark choices should be made, or a
completely different analysis framework should be considered.

\section{Case Study \#1: Simple vs.~Optimal Results}
\label{sec:case-study1}

When bidders are heterogeneous, with different valuation
distributions, the expected revenue-maximizing single-item auction can
be complex and highly dependent on the details of the distributions
(Section~\ref{sub:fail1}).  Are there simpler auctions that perform
almost as well?  Section~\ref{sub:chk} studies approximation
guarantees for the Vickrey auction supplemented with reserve prices.
Section~\ref{sub:tlevel} presents $t$-level auctions, which offer a
smooth trade-off between simplicity and optimality.
Section~\ref{sub:multi} discusses the state-of-the-art for
multi-item auctions.

\subsection{Vickrey with Reserves}\label{sub:chk}

Recall the single-item auction setting of Section~\ref{sub:fail1}.
There are $n$ bidders, with bidder $i$'s private valuation $\vali$
drawn independently from a distribution $\disti$ (with density
$\densi$) that is known to the seller.  We assume that
every distribution is regular, meaning that the corresponding virtual
valuation function  is nondecreasing.\footnote{Recall the virtual
  valuation function is given by 
$\vvi(\bidi) = \bidi - (1-\disti(\bidi))/\densi(\bidi)$.}
  The optimal auction
is a virtual welfare maximizer, and it computes a virtual
bid for each bidder, awards the item to the bidder with
highest positive virtual bid (if any), and charges the lowest winning
bid that the bidder could have made.  This auction depends in a
detailed way on the distributions $\dist_1,\ldots,\dist_n$.

Virtual welfare maximizers are a rich class of auctions, parameterized
by the virtual valuation functions $\vv_1,\ldots,\vv_n$.  Intuitively,
there is an infinite number of degrees of freedom in specifying such
an auction.  A natural and practically useful class of auctions with
far fewer parameters is that of reserve price-based auctions.  {\em Vickrey
  auctions with bidder-specific reserves} have only $n$ degrees of freedom,
the reserve prices $r_1,\ldots,r_n$.\footnote{The informal notion of
  ``degrees of freedom'' in an auction class can be made precise using
  concepts from statistical learning theory, such as the
  pseudodimension.  See \cite{MR15} for further details.}
Such an auction awards the item
to the highest bidder that meets its reserve, and charges the smallest
bid that would have won (the winning bidder's reserve price, or the
highest bid by a different bidder that clears its reserve, whichever
is larger).

Perhaps the most natural choice for bidder~$i$'s reserve price $r_i$
is the monopoly price for its distribution $\disti$
(Section~\ref{ss:iid}).  This choice guarantees a constant fraction of
the optimal expected revenue, where the constant is independent of the
number of bidders and the valuation distributions.
\begin{theorem}[Simple Versus Optimal Auctions]\label{t:simpleopt}
For all~$n \ge 1$ and regular distributions $\dist_1,\ldots,\dist_n$,
the expected revenue of an $n$-bidder single-item Vickrey auction with
monopoly reserve prices is at least 50\% of that of the optimal auction.
\end{theorem}
Thus, knowing a single statistic about each bidder's valuation
distribution (its monopoly price) already suffices for approximately
optimal expected revenue.\footnote{Even simpler are the Vickrey
  auctions with a single {\em anonymous} reserve price.  Anonymous
  reserve prices also suffice to extract a constant fraction of the
  optimal expected revenue, although the constant degrades to
  37\%~\citep{A+15}.}

Theorem~\ref{t:simpleopt} follows from \cite{CHK07} and \cite{HR09}.
It can also be derived from the ``prophet inequality'' of \cite{SC84};
see Chapter~4 of \cite{Har17} or Lecture~6 of \cite{Rou16}.

The guarantee of 50\% can be improved for many distributions.
It is tight in the worst case, however, even with only two
bidders and arbitrary, not necessarily monopoly, reserve
prices \citep{HR09}.  To improve the guarantee without making
additional assumptions, we must add complexity to the auction format.
The next section describes a principled way of doing so.

\subsection{$t$-Level Auctions and Simplicity-Optimality Trade-Offs}\label{sub:tlevel}
Virtual welfare maximizers are theoretically optimal but overly
complex.  Reserve-price-based auctions are reasonably simple but 
extract only
50\% of the optimal expected revenue in the worst case.
Can we interpolate between these two
extremes?  Can we quantify the trade off between simplicity and
optimality?  It's not clear how to make sense of this question 
without using the approximately optimal mechanism design paradigm.

\cite{MR15} proposed {\em $t$-level single-item auctions} for this
purpose.  Such an auction has $t$ parameters per bidder, which can be
viewed as an increasing sequence of $t$ reserve prices.  
Given a bid profile, the {\em
  level} of a bidder is defined as the number of its reserves that its
bid clears.  For example, if a bidder has three reserves~5,~7, and~9
and submits a bid of~8, then it has level~2.  

The allocation rule of a $t$-level auction is defined as follows.  If
every bidder has level~0, then the item remains unallocated.
Otherwise, the item is awarded to the bidder with the largest level,
with ties broken by bid.  That is, the winner is the highest bidder at
the top occupied level.  Since different bidders can have different
reserve prices, the winner need not be the highest bidder overall.  As
usual, the winning bidder pays the lowest bid at which it would
continue to win.  
1-level auctions are the same as Vickrey auctions with bidder-specific
reserves.

$t$-level auctions are naturally interpreted as discrete
approximations to virtual welfare maximizers.  Each level~$\ell$
corresponds to a constraint of the form ``If any bidder has level at
least $\ell$, do not sell to any bidder with level less than $\ell$.''
For every $\ell$, we can interpret bidders' $\ell$th 
reserve prices as the bidder values that map to some common virtual
value.  For example, $1$-level auctions treat all values below a
reserve price as having a negative virtual value, and above the
reserve use values as proxies for virtual values.  $2$-level
auctions use the second reserve to refine the virtual value
estimates, and so on.  With this interpretation, it is intuitively
clear that as $t\to\infty$, it is possible to estimate bidders'
virtual valuation functions and thus approximate Myerson's optimal
auction to arbitrary accuracy.  The next theorem is a quantitative
version of this intuition; for normalization purposes, it restricts
attention to distribution with support $[0,1]$.
The proof idea is to
``round'' an optimal auction to a $t$-level
auction without losing much expected revenue, using the reserve prices
to approximate each bidder's virtual value.

\begin{theorem}[\cite{MR15}]\label{t:mr15}
There is a constant $c > 0$ such that, for every number~$n$ of
bidders, $\eps > 0$, and valuation  distributions
$\dist_1,\ldots,\dist_n$ with support $[0,1]$, there is a $\tfrac{c}{\eps}$-level auction
with expected revenue within $\eps$ of optimal.
\end{theorem}

The guarantee in Theorem~\ref{t:mr15} translates to a relative
approximation of $1-\eps$ (with a different constant $c'$ in place of
$c$), except in the uninteresting case where the optimal expected
revenue is very close to~0.

\subsection{Multi-Parameter Problems}\label{sub:multi}

The approximation guarantees in Theorems~\ref{t:simpleopt}
and~\ref{t:mr15} hold more generally in most
single-parameter environments \citep{HR09,MR15}, almost at the level of
generality of Myerson's optimal auction theory \citep{Mye81}.

Multi-parameter problems like multi-item auctions, however, pose a
notorious challenge to optimal auction theory.  In most such settings,
there is no understanding of the optimal auction, other than being the
solution to an astronomically large linear program.  For an overview
of the solvable special cases, see \cite{DDT17} and the references
therein.

\cite{HR15} suggested studying the seemingly simple case of a single
buyer and multiple items, where the buyer has an additive valuation
and its values for different items are independent.  They documented
several troublesome and counterintuitive properties possessed by
optimal multi-item auctions, even in this restricted setting.

\cite{HN17} proposed using approximation to make progress on this
class of multi-item auction problems.  Passing to approximation can
bypass the challenge of characterizing the optimal auction.  The
reason is that the analyst can instead use an analytically tractable
upper bound on the optimal expected revenue, and prove that an auction
of interest captures a significant fraction of this upper bound.

\cite{HN17} focused on two simple mechanisms: selling items separately
(one price per item, with the buyer picking a utility-maximizing
bundle); and a take-it-or-leave-it offer for the bundle of all items.
They proved that, as the number of items grows large, neither
mechanism guarantees a constant
fraction of the optimal revenue.
In a significant advance, \cite{BILW14} proved
that, for every distribution over additive valuations with independent
item values, one of these two mechanisms extracts a constant fraction
of the optimal revenue.\footnote{Because the mechanism's expected
  revenue is compared to an upper
  bound on the optimal expected revenue,
  there are two sources of
  suboptimality: in the auction itself (due to revenue loss 
relative to an optimal auction), and in the analysis (due to slack
between the upper bound and the actual expected revenue of an optimal
  auction).  For this reason, the numerical value of the constant is
  not particularly satisfying when taken at face value.}
\cite{Yao15} extended this result to multiple buyers, and
\cite{RW15} to more general valuation distributions.

\section{Case Study \#2: Low-Dimensional Message Spaces}
\label{sec:case-study2}

In this section we switch gears and study 
the problem of allocating multiple items to bidders with private
valuations to maximize the social welfare.
We instantiate the approximately optimal mechanism design
paradigm to identify conditions on bidders' valuations that are
necessary and sufficient for the existence of simple combinatorial
auctions with near-optimal welfare at equilibrium.
The take-away from this section is that rich bidding spaces are
an essential feature of every good combinatorial auction when items
are complements, while simple auctions can perform well when bidders'
valuations are complement-free.

\subsection{Motivating Question}

Recall from
Section~\ref{sub:fail2}
the standard setup for allocating multiple items via a
combinatorial auction.  There are $n$ bidders and $m$ non-identical
items.  Each bidder has, in principle, a different private valuation
$\vali(S)$ for each bundle~$S$ of items it might receive.  
In this section, we assume
that the objective is to determine an allocation $S_1,\ldots,S_n$ that
maximizes the social welfare $\sum_{i=1}^n \vali(S_i)$.
The VCG mechanism is dominant-strategy incentive-compatible and
welfare-maximizing but, as a direct-revelation mechanism, it requires
an exorbitant number of bids ($2^m$) from each bidder.

In this case study, we apply the approximately optimal mechanism
design paradigm to study the following question.
\begin{itemize}
	
	\item [] {\em Does a near-optimal combinatorial auction require rich
		bidding spaces?}
	
\end{itemize}
Thus, as in the previous case study, we seek conditions
under which ``simple auctions'' can ``perform well.''
This time, our design space of ``simple auctions'' consists of mechanism
formats in which 
the dimension of every player's bid space is growing polynomially with
the number $m$ of items (say $m$ or $m^2$), rather than exponentially
with $m$ as in the VCG mechanism.

``Performing well'' means, as usual, achieving objective function value
(here, social welfare) close to that of a benchmark.
We use the {\em VCG benchmark}, meaning the welfare obtained by the
best arbitrarily complex mechanism (the VCG mechanism), which is
simply the maximum-possible social welfare.

This case study contributes to the debate about whether or not
package bidding 
is an important feature of combinatorial auctions, a topic
over which much blood and ink has been spilled over the past twenty
years.
We can identify auctions with no or limited packing bidding with
low-dimensional mechanisms, and those that support rich package
bidding with high-dimensional mechanisms.
With this interpretation, our results make precise the intuition that
flexible package bidding is crucial when items are complements, but
not otherwise.

\subsection{A Simple Auction: Selling Items Separately}

Our goal is to understand the power and limitations of the entire
design space of low-dimensional mechanisms.  
To make this goal more concrete, we begin by examining a
specific simple auction format.

The simplest way of selling multiple items is by selling each
separately.  Several specific auction formats implement this general
idea.  We analyze one such format, simultaneous first-price
auctions~\citep{B99}.  In this auction, each bidder submits
simultaneously one bid per item---only $m$ bidding parameters,
compared with its $2^m$ private parameters---and each item is sold
in parallel using a first-price auction.

When do we expect simultaneous first-price auctions to have reasonable
welfare at equilibrium?
Not always.  With general bidder valuations, and in particular when
items are complements, we might expect severe inefficiency due to
the ``exposure problem'' (e.g.,~\cite{Mil04}).  For example,
consider a bidder in an 
auction for wireless spectrum licenses that has large value for full
coverage of California but no value for partial coverage.
When items are sold separately, such a bidder has no vocabulary to
articulate its preferences, and runs the risk of obtaining a subset of
items for which it has no value, at a significant price.

Even when there are no complementarities amongst the items, we expect
inefficiency when items are sold separately (e.g.,~\cite{krishna}).
The first reason is 
``demand reduction,'' where a bidder pursues fewer items than it truly
wants, in order to obtain them at a cheaper price.
Second, if bidders' valuations are drawn independently from different
valuation distributions, then even with a single item,
Bayes-Nash equilibria are not always fully efficient.

\subsection{Valuation Classes}

\begin{figure}
	\begin{center}
		\includegraphics[scale=.5]{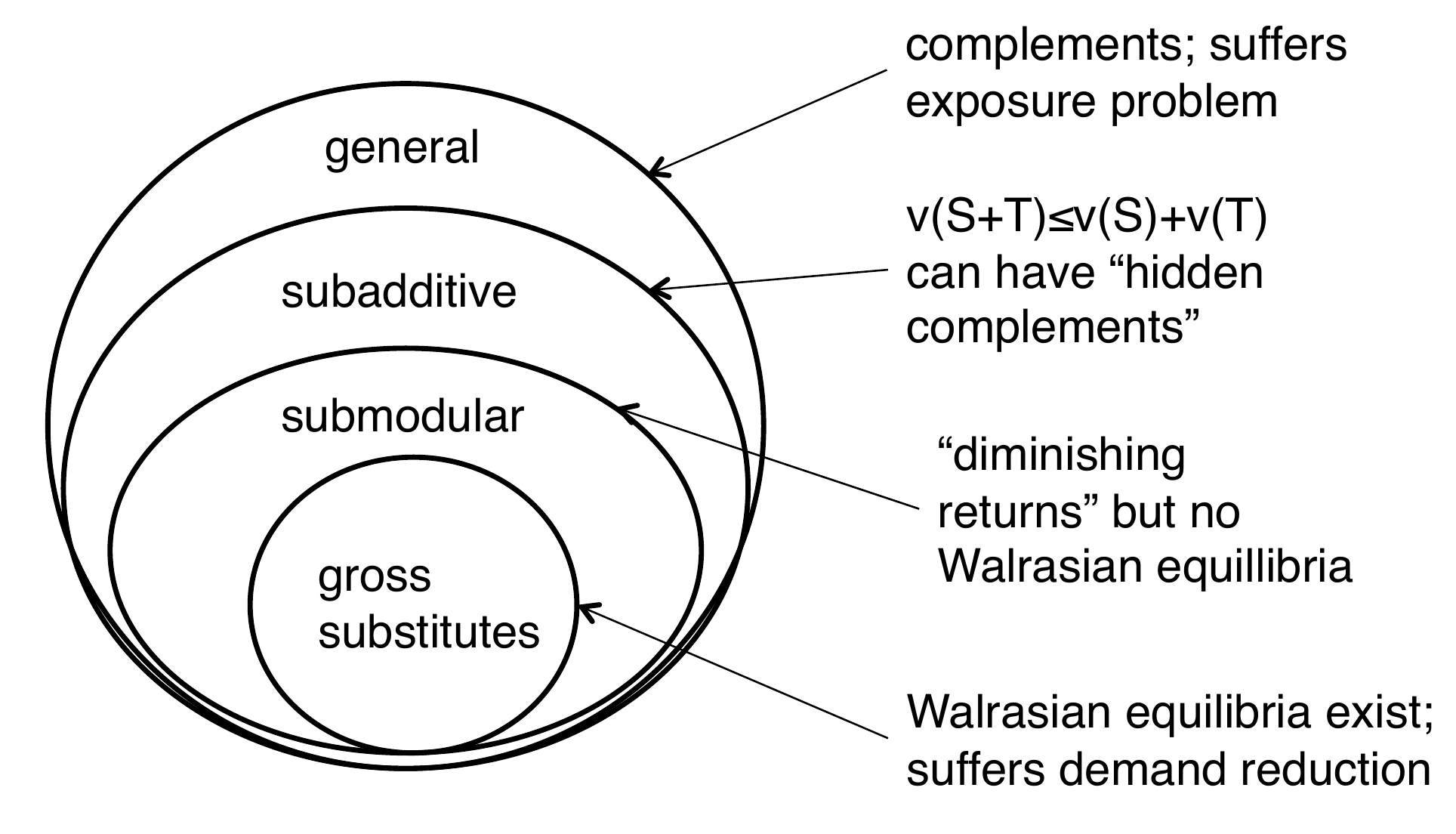}
		\caption{A hierarchy of valuation classes.}
		\label{f:vals}
	\end{center}
\end{figure}

Our discussion so far suggests that simultaneous first-price auctions
are unlikely to work well with general valuations, and suffer from
some degree of inefficiency even with simple bidder valuations.
To parameterize the performance of this auction format, we
introduce a hierarchy of bidder valuations (Figure~\ref{f:vals});
the literature also considers more fine-grained
hierarchies~\citep{feige,LLN06}.

The biggest set corresponds to general valuations, which can encode
complementarities among items.
The other three sets denote different notions of
``complement-free'' valuations.
In this survey, we focus on the most permissive of these, {\em
	subadditive valuations}.  Such a valuation $\vali$ is monotone
($\vali(T) \subseteq \vali(S)$ whenever $T \subseteq S$) and satisfies $\vali(S
\cup T) \le \vali(S) + \vali(T)$ for every pair $S,T$ of
bundles.
This class is significantly larger than the well-studied classes of
gross substitutes and submodular valuations.\footnote{Submodularity is 
	the set-theoretic analog of ``diminishing returns:'' $\vali(S \cup
	\{j\}) - \vali(S) \le \vali(T \cup \{j\}) - \vali(T)$ whenever $T
	\subseteq S$ and $j \notin S$.  The gross substitutes
        condition---which states that a bidder's demand for an
        item only  increases as the
	prices of other items rise---is strictly stronger and guarantees
	the existence of Walrasian equilibria~\citep{KelsoCrawford1982,GS99}.}
In particular, subadditive   valuations can have ``hidden
complements,'' with two items becoming complementary once
a third item is acquired, while submodular
valuations cannot~\citep{LLN06}.

\subsection{When Do Simultaneous First-Price Auctions Work Well?}\label{ss:s1a}

Our intuition about the performance of simultaneous first-price
auctions translates nicely into rigorous statements.  First, for
general valuations, selling items separately can be a disaster.
\begin{theorem}[\cite{H+11}]\label{t:h+11}
	With general bidder valuations, simultaneous first-price auctions can
	have mixed-strategy Nash equilibria with expected welfare arbitrarily
	smaller than the VCG benchmark.
\end{theorem}
For example, equilibria of simultaneous first-price auctions need not
obtain even 1\% of the maximum-possible welfare when there are
complementarities between many items.

On the positive side, even for the most permissive notion of
complement-free valuations---subadditive valuations---simultaneous
first-price auctions suffer only bounded welfare loss.
\begin{theorem}[\cite{FFGL13}]\label{t:ffgl13}
	If every bidder's valuation is drawn independently from a distribution
	over subadditive valuations, then the expected welfare obtained at
	every Bayes-Nash equilibrium of simultaneous first-price auctions
	is at least 50\% of the expected VCG benchmark value.
\end{theorem}
In Theorem~\ref{t:ffgl13}, the valuation distributions of different
bidders do not have to be identical, just independent.
The guarantee improves to roughly 63\% for the special case of
submodular bidder valuations~\citep{ST13}.

Taken together, Theorems~\ref{t:h+11} and~\ref{t:ffgl13} suggest that
simultaneous first-price auctions should work reasonably well if and
only if there are no complementarities among items.

\subsection{Negative Results}

We now return to the question of when simple mechanisms, meaning
mechanisms with low-dimensional bid spaces, can achieve non-trivial
welfare guarantees.  Section~\ref{ss:s1a} considered the special case of
simultaneous first-price auctions; here we consider the full design
space.

First, the poor performance of simultaneous first-price auctions 
with general bidder valuations is
not an artifact of the specific format: {\em every} simple mechanism
is vulnerable to arbitrarily large welfare loss when there are
complementarities among items.  This impossibility result argues
forcefully for a rich bidding language, such as flexible package
bidding, in such environments.
\begin{theorem}[\cite{R14}]\label{t:r14a}
	With general bidder valuations, no family of simple mechanisms
	guarantees equilibrium welfare at least a constant fraction of the
	VCG benchmark.
\end{theorem}
In Theorem~\ref{t:r14a}, the mechanism family is parameterized by the
number of items $m$; ``simple'' means that the number of dimensions in
each bidder's bid space is bounded above by some polynomial function of
$m$.  The theorem states that for every such family and constant $c >
0$, for all sufficiently large $m$, there is a valuation profile and
a full-information mixed Nash equilibrium of the mechanism with
expected welfare less 
than $c$ times the maximum possible.\footnote{Technically,
	Theorem~\ref{t:r14a} proves this statement for an $\eps$-approximate
	Nash equilibrium---meaning every player mixes only over strategies
	with expected utility within $\eps$ of a best response---where
	$\eps > 0$ can be made arbitrarily small.  The same comment applies
	to Theorem~\ref{t:r14b}.}
The proof of Theorem~\ref{t:r14a} builds on techniques from the field
of complexity theory, specifically communication complexity
\citep{KN97,Rou16b}.

We already know from Theorem~\ref{t:ffgl13} that, in contrast, simple
auctions can have non-trivial welfare guarantees with complement-free
bidder valuations.  Our final result states that no simple mechanism
outperforms simultaneous first-price auctions with these bidder valuations.
\begin{theorem}[\cite{R14}]\label{t:r14b}
	With subadditive bidder valuations, no family of simple mechanisms
	guarantees equilibrium welfare more than 50\% of the
	VCG benchmark.
\end{theorem}

\section{Case Study \#3: Algorithmic Mechanism Design}
\label{sec:case-study3}

In this section we address a third possible failure of optimal
mechanisms: excessive computation.  As in the previous section, we
study the problem of allocating resources to players with private
valuations to maximize the social welfare.

Section~\ref{sub:cc} introduces the theory of computational
complexity, which studies the amount of computational resources
necessary and sufficient to solve different computational problems.
Section~\ref{sub:motivating-ex} interprets this theory in the context
of the recent FCC Incentive Auction.  Section~\ref{sub:amd-main-Q}
states our design goals.  Sections~\ref{sub:amd-single-param}
and~\ref{sub:amd-multi-param} instantiate the approximately optimal
mechanism design paradigm in single- and multi-parameter settings,
respectively.  Our single-parameter example concerns allocating a
limited-capacity shared resource, and we'll see that ``greedy''
mechanisms often perform well.  Our main multi-parameter example is
multi-unit auctions, and we'll see how to modify the VCG mechanism,
with bounded loss of social welfare, so that it becomes
computationally tractable.

\subsection{Computational Complexity}\label{sub:cc}

The field of {\em computational complexity} analyzes the amount of
computational resources, such as the amount of time, required to solve
a computational problem.  Examples of computational problems include
sorting a given set of numbers, sequencing a given set of tasks,
computing a shortest path between a given origin and destination in a
network, and so on.
A positive result in this field takes the
form of a computationally efficient algorithm---an algorithm that
solves every instance of a problem in a reasonable amount of time.
The most common definition of ``reasonable'' is as a polynomial-time
algorithm: the running time
(i.e., number of elementary steps) performed by the algorithm grows as
a polynomial function of the size of the instance (e.g., the number of
tasks to be sequenced, or the number of vertices and edges in the
given network).  Equivalently, the input sizes that the algorithm can
solve in a fixed amount of time scales multiplicatively with
increasing computational power. 
An example of an inefficient algorithm is one that exhaustively searches
through an exponential number of possible solutions (cf., Section~\ref{sub:fail3}).
A standard textbook treatment of
computational complexity is \cite{sipser}; see also \cite{Rou10} for a
survey aimed at economists.

Unfortunately, many computational problems, including many that arise
in economics, are ``$NP$-hard.''  A formal definition of this term is
outside the scope of this survey, but the bottom line is that
$NP$-hard problems do not admit computationally efficient algorithms
under widely believed mathematical assumptions (specifically, the
``$P \neq NP$'' conjecture).

While the $NP$-hardness of a problem rules out any always-fast,
always-correct algorithm for the problem (assuming $P \neq NP$), it is
not a death sentence.  In some (but not all) applications, the
instances of an $NP$-hard problem relevant to practice are relatively
easy and can be solved in a reasonable amount of time.

\subsection{Computational Complexity in Practice: The FCC Incentive Auction}
\label{sub:motivating-ex}

The lessons of computational complexity show up frequently in the real
world.  For an example germane to this survey, consider the US FCC
Incentive Auction of 2016--17.  This auction consisted of two phases:
(i) freeing up a designated band of spectrum by buying it back from TV
broadcasters; and (ii) reselling the cleared spectrum to interested
companies \citep{Mil17}.  Two computational problems are closely
associated with the buying-back phase. The first is the problem of
checking whether a given set of broadcasters can stay on-air, i.e.,
can be feasibly repacked into the band of spectrum not designated for
sale to companies. In the second computational problem, given every
broadcaster's value for remaining on-air, the goal is to find the
subset of broadcasters with maximum total value (welfare) that can be
feasibly repacked.

While both computational problems are $NP$-hard, the problem of
checking feasibility can be reformulated as a \emph{satisfiability
  (SAT)} problem, for which effective SAT-solver software exists
\citep{NLMS17}.\footnote{An instance of the satisfiability problem
  is a logical formula in a specific format with a number of free Boolean
  variables; the question is whether it's possible to assign values to
  the free variables so that the formula is satisfied (i.e., is true).}
For the problem of welfare maximization, however, no effective heuristic is known.%
\footnote{\citet{Mil17}, Section 4.3: ``In the actual auction, Vickrey
  outcomes [...] cannot be computed at all.'' \label{foot:milgrom}} 
This demonstrates that computational complexity can be a true hurdle for mechanism design, forcing the designer to embrace an approximation approach, as was done in the FCC Incentive Auction. 

\subsection{Design Goals}
\label{sub:amd-main-Q}

For the rest of this section, our goal is to design a mechanism that:
(i) is dominant-strategy incentive-compatible, or DSIC (an important
requirement in our motivating example---the Incentive Auction should
be simple for broadcasters to participate in); (ii) is
welfare-maximizing, subject to feasibility; and (iii) runs in
polynomial time.  When the welfare maximization requirement involves
solving an $NP$-hard problem, properties (ii) and (iii) are
incompatible (even ignoring the DSIC requirement) and one of them must
be relaxed.  We consider relaxing~(ii) and settling for approximate
welfare-maximization.
A fundamental question in algorithmic game theory, first posed by
\cite{NR01}, is whether the DSIC requirement leads to further loss in
the approximation factor. In other words, is \emph{mechanism}
design fundamentally more difficult than \emph{algorithm}
design?\footnote{We focus on the case of DSIC mechanism design, 
where in general the answer is yes \citep{PSS08,Dob11,DV15,DV16,DSS18}.
The same question makes sense for Bayesian incentive-compatible
  (BIC) mechanisms, and for this version of the question, recent research has produced
  strong and general positive results \citep{HL15,HKM15,BH11,DHKN17}.}

\subsection{Approximation in Single-Parameter Settings}
\label{sub:amd-single-param}

We introduce a single-parameter abstraction of the packing scenario
described in our motivating example (Section
\ref{sub:motivating-ex}). There are $n$ players (broadcasters) with
single-parameter values $v_1,\dots,v_n$ for being chosen by the
mechanism (staying on-air). There is a feasibility constraint
$\mathcal{F}\subseteq 2^{[n]}$ over player sets, where
$A\in\mathcal{F}$ if and only if the player set $A$ can be feasibly
chosen (repacked).%
\footnote{We assume that $\mathcal{F}$ is downward-closed, i.e., if
  $A$ is feasible and $A'\subseteq A$ then $A'$ is also feasible.}
The auction outcome is the chosen (on-air) player set
$A^*\in\mathcal{F}$, and its welfare is $\sum_{i\in A^*}v_i$.

\cite{Mye81} characterizes the range of DSIC mechanisms in such
single-parameter settings as \emph{monotone} allocation rules coupled
with \emph{critical bid} payments (analogous to the second-price
payment).  The approximate mechanism design question therefore reduces in
these settings to approximation algorithm design, subject to the extra
condition of monotonicity.

\subsubsection{Example: Knapsack}

A knapsack constraint corresponds to a shared resource with limited capacity $W$. Every player has a publicly-known size $w_i\le W$ (e.g., the size of bandwidth it needs to stay on-air), and $A\in \mathcal{F}$ if and only if the set of players $A$ fits within the knapsack ($\sum_{i\in A} w_i \le W$).
The computational problem of finding a welfare-maximizing player set
among the sets that fit within the knapsack is $NP$-hard; assuming
$P\neq NP$, there is no polynomial-time algorithm that solves the
problem in general.

The classic algorithm of \cite{IK75} achieves the next best thing:
it guarantees $99\%$ of the optimal welfare in polynomial time.%
\footnote{In fact, the guarantee is $(100-c)\%$ of the welfare, where
  $c$ can be an arbitrarily small constant. Formally, for any
  precision parameter
  $\epsilon$, 
  the algorithm guarantees $(1-\epsilon) \cdot 100\%$ of the optimal welfare
  and runs in time polynomial in $n$ and $\frac{1}{\epsilon}$.
  Similar comments apply to uses of ``99\%'' later in this survey.}
This algorithm is not monotone, but \cite{BKV11} show how to tweak it
to get a monotone allocation rule without compromising on the
approximation factor. Together with Myerson's critical bid pricing,
this gives a mechanism that satisfies all three desiderata~(i)--(iii)
above (up to a tiny loss in welfare).

\subsubsection{A Simple Greedy-Based Mechanism}
\label{sub:greedy-knapsack}

Next we consider a mechanism for the knapsack problem that is based on
the ``greedy'' approach, which is remarkably simple to describe and
analyze.

\begin{mdframed}[style=offset,frametitle={DSIC Greedy-Based Mechanism for the Knapsack Problem}]
	\begin{enumerate}
		
\item 
Solicit bids $b_1,\dots,b_n$, and re-index the players so that $\frac{b_1}{w_1}\ge\cdots\ge  \frac{b_n}{w_n}$. 
		\item Choose the biggest prefix~$\{1,2,\ldots,i\}$
                  of players with total size at most~$W$.

\item Return either this greedy solution or the highest bidder,
whichever has a higher sum of bids.

\item Charge Myerson's critical bid prices.

\end{enumerate} 
\end{mdframed}

In effect, this mechanism greedily considers
players one at a time, ordered according to 
their ``bang-per-buck.''
(The second solution is needed
only to handle the case where there is a single bidder that is both
very big and also has a very high valuation.)
Holding all other bids fixed, by
bidding higher a player can only go from being unchosen to being
chosen by the mechanism. This monotonicity coupled with the pricing
rule ensures that the mechanism is DSIC.
It also has a non-trivial approximation guarantee:

\begin{theorem}[Folklore]
	\label{thm:greedy-knapsack}
The greedy-based mechanism for the
        knapsack problem runs in polynomial time and,
assuming truthful bids,
 achieves at least
        $50\%$ of the optimal welfare.
\end{theorem}

To see why, first imagine that the greedy prefix~$\{1,2,\ldots,i\}$ of
players filled up the knapsack exactly, with no space left over.  In
this case, this prefix is an optimal solution---the player ordering
ensures that every unit of space in the knapsack is used in the
most-valuable possible way.  The only issue is when the greedy prefix
leaves some of the knapsack unfilled, because the sum of the sizes of
the first $i$ players is less than $W$ and that of the first $i+1$
players is $W' > W$.  The prefix $\{1,2,\ldots,i+1\}$ would be optimal
for a knapsack with capacity $W'$, and hence is only
better-than-optimal for the smaller knapsack capacity~$W$.  Each of
the sets $\{1,2,\ldots,i\}$ and $\{i+1\}$ is a feasible solution with
the original capacity~$W$, so one of them captures at least 50\% of
the optimal welfare.  The greedy-based mechanism does at least as
well.\footnote{The guarantee of 50\% is tight in the worst case.  
Let $\eps > 0$ be arbitrarily small.
If
  there are 3 players with sizes
  $\frac{W}{2}+\epsilon,\frac{W}{2},\frac{W}{2}$ and valuations
  $\frac{W}{2}+2\epsilon,\frac{W}{2},\frac{W}{2}$, then the
  greedy-based mechanism chooses player 1 and achieves welfare
  $\frac{W}{2}+2\epsilon$, while the optimal solution (players 2 and
  3) has welfare $W$.}

\subsubsection{Performance of the Greedy Approach in Practice}
\label{sub:amd-bwca}

The FCC Incentive Auction is more complicated than a knapsack setting,
because its feasibility constraint $\mathcal{F}$ must take into account
not only sizes but also potential interferences among
geographically-close broadcasters. 
This results in a more difficult welfare-maximization problem, and
greedy approaches cannot guarantee any constant fraction of the
optimal welfare.
However,
when a greedy approach is applied in simulations, its empirical
performance is excellent, achieving $95\%$ of the optimum on average
\citep{NLMS17}.\footnote{The problem
instances considered by \cite{NLMS17} were kept small enough that the exponential-time
computation of the benchmark (the optimal welfare) could be carried out in a
reasonable amount of time.}

What characteristics of ``typical instances'' make them easier to
approximate that arbitrary instances?  
Approximation helps us identify relevant parameters
that govern the difficulty of welfare
maximization. \cite{Mil17} points to the distance of $\mathcal{F}$
from a matroid (see, e.g., \cite{oxley}) as one such parameter. 
Back in knapsack settings, item sizes are a crucial parameter:
if the size of every item is
at most $\alpha\%$ of the knapsack capacity (say $5\%$
or $10\%$), then the greedy-based approach guarantees 
welfare within $(100-\alpha)\%$
of optimal, even in the worst case (see, e.g.,
\cite{DRT17}). 
These examples illustrate how stronger worst-case guarantees are often
possible under stronger assumptions about the instances of interest.

\subsection{Approximation in Multi-Parameter Settings}
\label{sub:amd-multi-param}

As we have seen, in single-parameter settings there is a successful
paradigm for designing computationally efficient mechanisms with good
approximation guarantees: (i) Characterize the design space of
implementable algorithms (i.e., monotone allocation rules); (ii)
Optimize over this design space (i.e., find the best
computationally efficient monotone algorithm), or use a simple algorithm
from the design space (like a greedy algorithm).
This paradigm has had limited success in multi-parameter settings. The
reason is that the characterization of implementable multi-parameter
allocation rules (the ``cyclic monotonicity'' condition of
\cite{Roc87}) is quite inconvenient to work with.

An alternative idea focuses on the VCG mechanism, which can be seen as
an ingenious way to transform a welfare-maximizing algorithm into a
DSIC mechanism. Can this method be extended to
\emph{approximately} welfare-maximizing algorithms? 
Unfortunately, plugging an arbitrary approximation algorithm into the
VCG mechanism does not generally preserve incentive-compatibility
\citep{NR06}.  
A more specific approach is to
first commit to a restricted range of outcomes (prior to receiving
players' bids), and then, given players' bids, to run the VCG mechanism
with respect to the restricted range.
The hope is to define a range small or
well-structured enough to enable computationally-efficient welfare
maximization over it, yet large enough to contain a near-optimal
outcome for every valuation profile.  The resulting DSIC
mechanisms are called \emph{maximum-in-range (MIR)} mechanisms.

For example, consider a multi-unit auction setting with $n$ bidders
and $m$ homogeneous items (called units).  We do not assume that
bidders have decreasing marginal values.
Assume that $m$ is much
bigger than $n$ ($m=2^n$, say), and that the goal is to compute an
approximately welfare-maximizing allocation in time polynomial in $n$
and $\log_2 m$.
\footnote{Exact welfare-maximization is provably
  impossible under this time constraint, even ignoring
  incentive-compatibility.  If the running time need only be polynomial in
  $m$ rather than $\log_2 m$, then the VCG mechanism can be
  implemented efficiently using an algorithmic technique called
  dynamic programming.}
Such a computation cannot even take the time to examine a
bidder's valuation for each of the~$m$ possible allocations to it.

One simple maximal-in-range solution is to commit to selling units
only in multiples of $m/n^2$---equivalently, to bundle the units into
$n^2$ blocks of equal size---and then implement the VCG mechanism for
this restricted range using dynamic programming.  The resulting
mechanism uses computation polynomial in $n$ and $\log_2 m$, and is
guaranteed to produce an allocation with near-optimal welfare.

\begin{theorem}[\cite{DN10}]
	\label{thm:MIR-multi-unit}
The maximum-in-range multi-unit auction above
runs in time polynomial in $n$ and $\log m$ and,
assuming truthful bids, 
 achieves at least
$50\%$ of the optimal welfare.
\end{theorem}

Can we do better?  Not with maximum-in-range mechanisms:
\cite{DN10} prove that no such mechanism
can guarantee  more than $50\%$ of the optimal welfare.
What about with a more general class of mechanisms?

A \emph{randomized} mechanism outputs a distribution over allocations
and payments. It is \emph{truthful (in expectation)} if for any
valuation reports of the other players, the expected utility of a
player is maximized by reporting truthfully, where the expectation is
over the randomization internal to the mechanism.  The
maximum-in-range approach can be immediately generalized to randomized
mechanisms: A \emph{maximum-in-distributional-range (MIDR)} mechanism
pre-decides on a range of distributions over allocations; based on the
reported valuations, it chooses the distribution from the range that
induces the maximum expected welfare.  
MIDR mechanims are sufficiently powerful to obtain almost optimal
welfare in multi-unit auctions:

\begin{theorem}[\cite{DD13}]
	There exists a maximum-in-distri\-butional-range mechanism for
        homogeneous items that runs in time polynomial in $n$ and
        $\log m$ and achieves in expectation at least $99\%$ of the optimal welfare.
\end{theorem}

The maximum-in-distributional-range approach is useful 
also in multi-item auctions with non-identical items.
Circling back to FCC spectrum auctions, a reasonable model for how
potential spectrum buyers value sets of channels is the class of
\emph{coverage} valuations, which assign value to every channel set
according to the population it covers. There is a computationally
efficient MIDR 
mechanism based on convex programming that achieves roughly $63\%$ of the
optimal welfare for bidders with coverage valuations \citep{DRY16}.

\section{Discussion and Alternatives to Approximation}
\label{sec:alternatives}

\subsection{A Second Look at Approximation}

Fundamentally, the goal of the approximately optimal mechanism design
framework is to make comparisons between competing mechanisms for a
problem and identify a ``best-in-class'' mechanism.
Even with a given objective function, it is not always
obvious how to compare two different mechanisms: the first will
generally perform better in some cases (e.g., for some valuation
profiles, or some prior distributions) and the second in other cases.
This issue does not come up in classical welfare-maximization
settings: because the VCG mechanism maximizes the social welfare ex
post, it is better than every competing mechanism (with respect to the
social welfare objective) irrespective of the valuation profile.  It
also does not arise in the traditional optimal auction setting: once
the prior distribution is fixed, competing auctions can be
unequivocally ranked according to expected revenue.

In many of the canonical applications of approximately optimal
mechanism design, including the ones described in this survey, the
theoretically optimal mechanism serves only as a benchmark.  The
performance of a mechanism of interest (like a ``simple'' mechanism)
is assessed through a relative comparison to this benchmark.  
A given mechanism
generally approximates the benchmark better in some cases than others,
and two different mechanisms generally have incomparable performance,
which each one superior to the other in some cases.  In approximately
optimal mechanism design, the performance of a mechanism is usually
summarized with a single number, by taking the worst-case (i.e.,
minimum) performance guarantee achieved in a setting of interest.  For
example, the 50\% guarantee in Theorem~\ref{t:simpleopt} holds in the
worst case over all single-item auction settings with regular
valuation distributions, the 50\% guarantee in Theorem~\ref{t:ffgl13}
holds in the worst case over all subadditive valuation distributions,
and the 50\% guarantee in Theorem~\ref{thm:greedy-knapsack} holds in
the worst case over all valuation profiles.\footnote{All three results
  are also tight in a worst-case sense, meaning there exists a setting
  of interest in which the mechanism's approximation guarantee
  achieves the worst-case bound.}  With this single number attached to
every mechanism, there is an obvious way to compare them, according to
their worst-case approximation guarantees.
This approach seems to enable results that would be hard or impossible
to establish by other means, such as the quantification of
simplicity-optimality trade-offs in Theorem~\ref{t:mr15} and the
optimality of simultaneous first-price auctions among low-dimensional mechanisms (Theorems~\ref{t:ffgl13} and~\ref{t:r14b}).%
\footnote{Making the right modeling choices can be key to obtaining
  meaningful qualitative insights.  One choice is the settings of
  interest---for example, the restriction to regular distributions in
  Theorem~\ref{t:simpleopt} or subadditive valuations in
  Theorem~\ref{t:ffgl13}.  The choice of the benchmark can also
be  important.  For example, should the performance of a simple DSIC
  mechanism be compared to that of the best arbitrarily complex DSIC
  mechanism, or the best BIC mechanism?  Of course, the more
  permissive the benchmark, the harder it is to approximate it.}

Instead of taking the worst case over all settings of interest, why
not take a Bayesian approach and impose a prior distribution over
these settings?
For example, in the single-item context of Theorem~\ref{t:simpleopt},
we could assume that the seller has a ``higher-order belief'' in the
form of a distribution over valuation distributions.  But as pointed
out by \cite{Seg03}, ``the dependence on the seller's prior is simply
pushed to a higher level.''  Complexity and detail-dependence---the
very disadvantages we were trying to avoid---creep back in, rendering
this approach ineffective for our purposes.

In the remainder of this section, we discuss other approaches from the
literature that aim to reveal useful mechanism design techniques or
understand simplicity-optimality tradeoffs.  Each of these approaches
has both merits and downsides, and each is an important tool in the
market designer's toolbox.

\subsection{Alternative I: Robust Max-Min Optimality}
\label{sub:robust}

Robust mechanism design is a rising paradigm in economics, surveyed in
this issue by \cite{Car18}.  It has roots in operations research and
robust optimization, where uncertainty about parameters is represented
by deterministic ``uncertainty sets,'' and optimization is in the
max-min sense over these sets.\footnote{See the work of \cite{BB14} for an
application of robust optimization to auctions.}  In recent years,
robust mechanism design has offered justifications for the ubiquity of
several popular auction and contract formats, often capturing the
common wisdom about what makes them so widely embraced in practice.

Robust mechanism design shares some of the ``worst-case'' or
``max-min'' flavor of approximately optimal mechanism design, but
it strives for max-min optimality rather than
the optimal approximation of a benchmark.
For concreteness, consider robustness against detailed knowledge of an
exponential-sized joint distribution, representing the distribution of
values for bundles of goods. Following \cite{Car17}, assume instead
knowledge only of the \emph{marginal} distributions of values of
individual goods. The paradigm of robust mechanism design replaces the
original model, in which every market instance corresponds to a joint
distribution, with a new model, in which every market instance
corresponds to a set of marginals. The objective of maximizing the
expected revenue for each instance is then replaced with a max-min
objective: we measure a mechanism's performance on a ``partial''
instance (i.e., marginals) by its performance on the worst-case
``full'' instance (i.e., joint distribution) that is compatible with
the partial instance. The result is a new problem formulation with new
instances (partial) and objective (max-min) to maximize. In this new
model, the optimal mechanism is well defined---it is the mechanism
that maximizes the max-min objective for every partial instance.  A
successful outcome of this modeling approach could be the novel
justification of a widely-used mechanism format as the robustly optimal
solution, or the identification of a new and potentially useful
robustly optimal mechanism.

Robust mechanism design makes weaker informational assumptions than
its classical counterpart, thus tackling head-on the problem of
excessive detail-dependence.
When the primary criticism of the optimal mechanism stems from sources
like communication or computational complexity rather than
detail-dependence, however, it is not immediately obvious how to apply
to robust mechanism design perspective.
In this case, simple mechanisms are desirable due to their practical
implementability, not their robustness to details of the environment
per se.

\subsection{Alternative II: Asymptotic Optimality (Large Market) Results}
\label{sub:large-markets}

In asymptotic analysis, the performance of a simple mechanism is
measured as the size of the market (number of players) goes to
infinity. The hope is to establish that the mechanism becomes optimal
in the limit, despite its simplicity. Such a result would imply that
as long as the market is sufficiently ``large,'' we can combine the
best of both worlds (simplicity and almost-optimality).
When successful, the asymptotic approach gives a new
sense in which simple mechanisms can be very close to optimal.

The intuition behind ``large market results'' is as follows: As more players participate in a resource-allocation mechanism, 
the actions of a single player have an \emph{increasingly negligible}
effect on the prices and outcome of the mechanism (under certain
conditions).\footnote{See \cite{RP76} for an early example of
  quantifying this intuition.}
This makes the players more homogeneous from the mechanism's perspective, and their behavior easier to incentivize; 
it then becomes possible even for simple mechanisms to optimize economic objectives such as revenue and welfare.
Put differently, worst-case bounds on inefficiency can
be overly pessimistic when determined by pathological ``small market''
examples, and this issue goes away in larger markets. 

A central example of the asymptotic optimality approach is the work of
\cite{Swi01}, who studies simple auctions for asymmetric players
competing over a single good (with multiple units).  Whereas in small
markets such auctions can have very inefficient equilibria,
\cite{Swi01} shows that under certain conditions of ``noisy'' demand
or supply, the equilibria become arbitrarily close to
welfare-maximizing as the markets grow large.  Roughly, the role of
the noisy demand/supply assumption is to provide enough randomness to
rule out pathological examples with inefficient equilibria that
persist even in large markets.  \cite{FIL+16} extend these results to
multiple different goods, as well as to additional simple auction
formats.  They also combine the large market approach with the
approximation paradigm; in one of the settings they study, equilibrium
inefficiency decreases but does not vanish in the limit.
\cite{Seg03} establishes a large market
result for revenue rather than welfare, showing how the revenue
performance of detail-free auctions can converge quickly to that of
optimal detail-dependent ones; see also \cite{BV03}, \cite{G+06}, and
\cite{N03} for related results.
Other large market results exist in
the context of resource allocation without money, some showing that
simple mechanisms increase in efficiency as the market grows large
(e.g., \cite{CK10}), others that they gain good properties like
truthfulness or stability (e.g., \cite{CKK18}).

There are two main downsides to the asymptotic large market approach:
(i) no guarantees for small- to medium-sized markets,
which are common in many modern applications (e.g., many keyword
advertising auctions); and (ii) in some cases, reliance on sufficient
randomness in the market.

\subsection{Alternative III: Resource Augmentation}

\cite{BK96} first introduced the approach of resource augmentation as
an alternative to complex optimal mechanisms. In their seminal
auctions vs.\ negotiations result, they show that in a symmetric
single-item auction setting with a common regular valuation
distribution,
running a standard Vickrey auction with one extra (i.i.d.) bidder
earns at least as much expected revenue as a (distribution-dependent)
optimal auction without the extra bidder.  We can view the
players as ``resources,'' in the sense that their competition is what
drives up prices and generates high revenue. Adding a player can 
be viewed as an ``augmentation'' of these resources.\footnote{The
  phrase ``resource augmentation'' was originally coined by \cite{KP00}
  in the context of algorithm analysis, inspired in part by \cite{ST85}.}
While the intuition of
more competition leading to more revenue is clear, it is not a priori
apparent by how much the competition needs to be increased for a
simple mechanism to outperform the optimal one (and whether this is at
all possible for a particular simple mechanism).

\cite{RTY17} generalize the result of \cite{BK96} to multi-item auctions, and also make two connections to approximation: first, by combining resource augmentation with approximation to limit the required amount of extra resources; and second, by establishing a framework for transforming a resource augmentation guarantee into an approximation one. Related approximation guarantees were established by \cite{DHKN11}. 
The work of \cite{EFF+17} further generalizes the approach of augmenting competition by applying it to more restrictive benchmarks and challenging revenue maximization settings. \cite{FFR18} combine resource augmentation with an approximation guarantee of $99\%$, and \cite{LP18} apply this approach to dynamic mechanisms. 

One downside of resource augmentation is that comparing the augmented
mechanism to the optimal one with no augmentation is in some sense
``unfair,'' like comparing apples to oranges. On the other hand, this
approach enables a direct comparison between the cost of resource
augmentation and the cost of mechanism complexity, which is important
for making an informed choice between a simple mechanism and its
complex counterpart.  Another limitation of the competition
enhancement approach 
is that it typically assumes that 
players are i.i.d.~and regular; for possible solutions to this issue,
see \cite{HR09} and \cite{SS13}. Finally, resource augmentation has been applied
so far mainly to revenue-maximization problems, although it has
recently been adopted for other domains as well \citep{AMS18}.

\section{Conclusion}
\label{sec:summary}

The main message of this survey is that approximation is useful for
achieving \emph{qualitative insights} on mechanism design in
\emph{complex settings}.  Section \ref{sub:summary-cases} 
summarizes briefly our supporting evidence for this statement, and the
takeaways from our three case studies.  Complexity is quickly becoming
the norm, and even the defining feature, in many important economic
settings \citep{Nis17}. Many modern transactions take place
within complex market environments---ridesharing
platforms, crowdsourcing marketplaces, 
and so on.
This suggests that, like other techniques for dealing
with complexity, the approximation paradigm will only increase in
utility in the coming years.

At present, however, approximation is possibly the most polarizing
topic in debate among computer scientists and economists working on
mechanism design. Economic theorists have largely ignored this
paradigm, passing on the opportunity to add to their arsenal a
well-developed and deep mathematical toolbox.\footnote{See
  \cite{ABM16} for a recent exception.}  For their part, computer
scientists have arguably been guilty of devoting disproportional
effort to small improvements in approximation factors, often at the
expense of useful qualitative insights.  They have also been accused
of viewing every problem as a nail to which the approximation hammer
should be applied.

We postulate that many if not all of these issues are caused by
an overly literal interpretation of approximation factors, as detailed
in Section \ref{sec:approx-paradigm}.  In this sense, the debate on
approximation has greatly advanced the research community's
understanding of the meaning behind approximation guarantees in
mechanism design.  It goes without saying that a guarantee of (say)
$50\%$ does not in itself justify the use of a mechanism; but in the
context of a challenging area of mechanism design, in which there is
no useful characterization of the optimal mechanism and no explanation
as to why certain mechanisms are observed in practice, the same
guarantee of a $50\%$-approximation can become meaningful and
enlightening.

As \cite{Car18} notes, the culture of economic theory is becoming
gradually more pluralistic; we believe this is an excellent time for
economists to take another look at approximation---at the very least,
as a useful complement to the widely-accepted approaches in Section
\ref{sec:alternatives}.  In the best-case scenario, approximation
could become a leading example of the kind of gains that stem from
interdisciplinary research.

\subsection{Summary of Case Studies}
\label{sub:summary-cases}

We briefly summarize the benefits of applying the approximation
paradigm to address the following three complexity barriers in
mechanism design: (1) opaque and detail-dependent mechanisms;
(2) unreasonable communication requirements from participants; and (3)
prohibitive computational complexity.

Section~\ref{sec:case-study1} considered barrier (1).
In single-parameter environments, the pursuit of approximation
guarantees guided us to a parameterized family of mechanisms 
that runs the gamut from complex optimal mechanisms to simple
mechanisms with constant-factor approximation guarantees.
In multi-parameter environments, the 
approximation approach led to a relatively simple mechanism---selling
items separately or as a single bundle, whichever is better---that
provably extracts a constant fraction of the optimal expected revenue
in certain multi-item auction settings.

Section~\ref{sec:case-study2} focused on barrier (2).
Here, the approximation perspective confirms that the
simple common method of selling items separately has excellent
welfare-performance guarantees, as long as bidders' preferences do not
include strong complementarities between items.
This mirrors the conventional wisdom among both theoreticians and
practitioners in multi-item auction design.

Section~\ref{sec:case-study3} addressed barrier (3), and showed that a
natural greedy approach to sharing a limited-capacity resource is
near-optimal in theory, and significantly exceeds expectations in
practice. Meanwhile, in certain multi-item auctions, generalizations
of the VCG mechanism achieve near-optimal welfare.

In all three applications, traditional economic tools targeted at
characterizing exact and optimal solutions appear inadequate to
achieve similar results.

\subsection{Directions for Future Research}

We highlight three directions for further research on approximation
and mechanism design: new applications for the approximation paradigm;
improved understanding of the relationships between different notions of
complexity (and the corresponding approximation guarantees); and
narrowing the gaps between worst-case analysis and the ``typical
cases'' relevant to practice.  We believe that more research in these
directions is necessary to expand the reach of the theory and improve
its coherence and applicability.  Many additional questions arise in
relation to existing application areas of the approximation paradigm,
like those in our case studies.

\begin{enumerate}
	
\item {\bf New frontiers for approximation.} Cutting-edge economic
  theory and new applications can inspire new ways for the 
  approximation paradigm to contribute.  One source of new
  frontiers is market design in practice. For
  example, as part of the design of the FCC Incentive Auction,
  \cite{MS17} developed a \emph{reverse greedy heuristic} and analyzed
  its strong incentive properties. The approximation paradigm can be
  used to study formally its welfare guarantees
\citep{DGR17,GMR17}.
	
Another source for new opportunities is
classical areas of economics in which complexity matters, and perhaps has
been studied using one the approaches in
Section~\ref{sec:alternatives}, but to which the
approximation paradigm has not yet been applied.
For example, \emph{contract design} is a major success story of robust
mechanism design \citep{Car15}, and only very recently has the
approximation lens been applied to it 
\citep{DRT18}. 
Similarly, the large market approach (Section~\ref{sub:large-markets})
has been successful in achieving fair allocations in conjunction with
efficiency and strategyproofness \citep{CK10}. To relax the large
market requirement, researchers are beginning to explore different
notions of \emph{approximate} fairness \citep{Bud11,CKM+18}.  

Ideas can also flow in the opposite direction.
For example, we used the approximation paradigm in
Section~\ref{sec:case-study1} to formalize simplicity-optimality
trade-offs.  Could such trade-offs also be formulated using one of the
alternative approaches in Section~\ref{sec:alternatives}?
For example, are there settings where the robustly optimal mechanism
(in the sense of Section~\ref{sub:robust})
becomes gradually more complex as more details
of the environment are revealed?
	
\item {\bf Relations among different complexity measures.}  We have
  demonstrated how the approximation paradigm helps tackle different
  types of complexity.
These have largely been studied in isolation, but
researchers are now aspiring to
a more holistic and comprehensive understanding of mechanism
  design complexity. For example, in Section \ref{sec:case-study1} we
  discussed revenue approximation guarantees for classes of
  mechanisms with low information requirements. Do any of our
  conclusions change dramatically if we also enforce computational
  tractability? See \cite{GN17} for a recent example of work in this
  direction.
	
  Similarly, Section \ref{sec:case-study2} discussed equilibrium welfare
guarantees for auctions with low
  communication requirements.
What if we relax the assumption of convergence to equilibrium, perhaps
assuming instead some form of natural dynamics that requires less
computation?
See \cite{RST17} for a more
  detailed discussion of this point.

  A final example is the recent work of \cite{Dob16}, who proved a
  formal relationship between different measures of mechanism
  complexity---some measures related to the mechanism's format (when
  viewed as a menu of prices, e.g., the number of such prices), and
  some to its required resources (communication or
  computation). Extending such results to additional complexity
  measures would contribute to a more unified theory of
  mechanism complexity, and thus also of approximately optimal
  mechanisms.
	
\item {\bf Realistic (rather than pessimistic) models of complexity.}
  In Section \ref{sub:amd-bwca} we saw an example of the gap between
  worst-case approximation guarantees for greedy-based mechanisms and
  their (much better) performance in practice.  The ``beyond
  worst-case analysis'' research agenda in computer science advocates
  sharper analysis of algorithms to capture their true behavior,
  usually by focusing attention on a subset of the ``most relevant''
  inputs.  The same agenda is relevant for the analysis of
  mechanisms---we wish to develop models
  that explain when and why the empirical performance of simple
  mechanisms 
  significantly exceeds their worst-case approximation guarantees. 
See \cite{PSW18}
for a recent effort in this direction.

\cite{EGW11} and \cite{BBEW14} approached this issue from the
perspective of revealed preference and showed that, in certain
settings,
every rationalizable set of choice data is in fact consistent with an
easy optimization problem.
It would be interesting to extend this approach to other settings,
such as quasi-linear markets.

\end{enumerate}

\section*{Acknowledgments}

The first author is supported in part by NSF Award CCF-1813188 and a
Guggenheim Fellowship, and performed this work in part while visiting
the London School of Economics.
The second author is supported in part by the Israel Science Foundation Grant No.~336/18.

\end{document}